\documentclass[10pt,conference, compsocconf]{IEEEtran}
\pdfoutput=1

\usepackage{epsfig}
\usepackage{graphicx}
\usepackage{amssymb}
\usepackage[noend]{algorithmic}
\usepackage{algorithm}
\usepackage{multirow}
\usepackage{subfigure}
\usepackage[cmex10]{amsmath}
\usepackage{balance}
\usepackage{sidecap}

\begin{document}
%

\title{Comparative Performance Analysis of Intel Xeon Phi, GPU, and CPU}
\author{
George Teodoro$^1$, Tahsin Kurc$^{2,3}$, Jun Kong$^4$, Lee Cooper$^4$, and Joel Saltz$^2$\\
\IEEEauthorblockA{$^1$Department of Computer Science, University of Bras\'ilia, Bras\'ilia, DF, Brazil\\
$^2$Department of Biomedical Informatics, Stony Brook University, Stony Brook, NY, USA \\
$^3$Scientific Data Group, Oak Ridge National Laboratory, Oak Ridge, TN, USA\\
$^4$Department of Biomedical Informatics, Emory University, Atlanta, GA, USA}
\vspace{-4mm}
}

\maketitle
\vspace*{-3ex}
\begin{abstract}
We investigate and characterize the performance of an important
class of operations on GPUs and Many Integrated Core (MIC) architectures. Our
work is motivated by applications that analyze low-dimensional spatial datasets
captured by high resolution sensors, such as image datasets obtained from whole
slide tissue specimens using microscopy image scanners. We identify the data access
and computation patterns of operations in object segmentation and feature
computation categories. We systematically implement and evaluate the
performance of these core operations on modern CPUs, GPUs, and MIC systems for 
a microscopy image analysis application. Our results show that (1) 
the data access pattern and parallelization strategy employed by the operations
strongly affect their performance. While the performance on a MIC of operations that 
perform regular data
access is comparable or sometimes better than that on a GPU; (2) GPUs are
significantly more efficient than MICs for operations and algorithms that
irregularly access data. This is a result of the low performance of the latter
when it comes to random data access; (3) adequate
coordinated execution on MICs and CPUs using a performance aware task
scheduling strategy improves about 1.29$\times$ over a first-come-first-served 
strategy. The example application attained an efficiency of 84\% in an
execution with of 192 nodes (3072 CPU cores and 192 MICs).
\end{abstract}


\IEEEpeerreviewmaketitle

\section{Introduction} \label{sec:intro}
Scientific computing using co-processors (accelerators) has gained popularity in 
recent years. The utility of graphics processing units (GPUs), for example, 
has been demonstrated and evaluated in several application 
domains~\cite{cudaapps}. As a result, hybrid systems that combine multi-core 
CPUs with one or more co-processors of the same or different types are being
more widely employed to speed up expensive computations. 
The architectures and programming models of co-processors may differ from CPUs 
and vary among different co-processor types. This heterogeneity leads to challenging 
problems in implementing application operations and obtaining the best 
performance. The performance of an application operation will depend on the operation's 
data access and processing patterns, and may vary widely from one co-processor 
to another.  Understanding the performance characteristics of classes of operations 
can help in designing more efficient applications, choosing the appropriate co-processor 
for an application, and developing more effective task scheduling 
and mapping strategies. 

In this paper, we investigate and characterize the performance of an important
class of operations on GPUs and Intel Xeon Phi Many Integrated Core (MIC)
architectures.  Our
primary motivating application is digital Pathology involving the analysis of
images obtained from whole slide tissue specimens using microscopy image
scanners.  Digital Pathology is a relatively new application domain and imaging
modality compared to magnetic resonance imaging and computed tomography.
Nevertheless, it is an important application domain because investigation
of disease morphology at the cellular and sub-cellular level can reveal
important clues about disease mechanisms that are not possible to capture by other
imaging modalities. Analysis of a whole slide tissue image is both data and
computation intensive because of the complexity of analysis operations and data
sizes -- a three-channel color image captured by a state-of-the-art scanner can
reach 100K$\times$100K pixels in resolution. Compounding this problem is the
fact that modern scanners are capable of capturing images rapidly, facilitating
research studies to gather thousands of images. Moreover, an image dataset may be
analyzed multiple times to look for different features or quantify sensitivity
of analysis to input parameters. 

Although the microscopy image analysis is our main motivating
application, we expect that our findings in this work will be applicable 
in other applications. Microscopy image analysis belongs to a class of 
applications that analyze low-dimensional spatial datasets captured by 
high resolution sensors. This class of applications  
include those that process data from satellites and ground-based sensors in weather and
climate modeling; analyze satellite data in large scale biomass monitoring and change
analyses; analyze seismic surveys in subsurface and reservoir characterization; and
process wide field survey telescope datasets in
astronomy~\cite{508406,Chandola:2011:SGP:2024035.2024041,vatsavai-2011,ParasharMLADW05}.
Datasets in these applications are generally represented in
low-dimensional spaces (typically a 2D or 3D coordinate system); typical 
data processing steps include
identification or segmentation of objects of interest and characterization of the 
objects (and data subsets) via a set of features. 
Table~\ref{tab:op-categories} lists the categories of common operations in these application 
domains and presents examples in microscopy image analysis.  
Operations in these categories produce different levels of data products that
can be consumed by client applications. For example, a client application may
request only a subset of satellite imagery data covering the east coast of the
US. Operations from different categories can be chained to form analysis workflows to create
other types of data products. The data access and processing patterns in these
operation categories range from local and regular to irregular and global
access to data. Local data access patterns correspond to accesses to a single
data element or data elements within a small neighborhood in a spatial and
temporal region (e.g., data cleaning and low-level transformations). Regular
access patterns involve sweeps over data elements, while irregular accesses may
involve accesses to data elements in a random manner (e.g., certain types of
object classification algorithms, morphological reconstruction operations in
object segmentation). Some data access patterns may involve generalized
reductions and comparisons (e.g., aggregation) and indexed access (e.g.,
queries for data subsetting and change quantification). 
\begin{table}
\begin{center}
\caption{Operation Categories} 
\begin{footnotesize}
\begin{tabular}{|p{0.15\textwidth}|p{0.27\textwidth}|} 
\hline
Operation Category 	& Microscopy Image Analysis \\ \hline \hline 
Data Cleaning and Low Level Transformations & 
Color normalization. Thresholding of pixel and regional gray scale values. \\ \hline
Object Segmentation & 
Segmentation of nuclei and cells. \\ \hline
Feature Computation & 
Compute texture and shape features for each cell. \\ \hline
Aggregation & 
Aggregation of object features for per image features. \\ \hline 
Classification & 
Clustering of nuclei and/or images into groups. \\ \hline
Spatio-temporal Mapping and Registration & 
Deformable registration of images to anatomical atlas. \\ \hline 
Data Subsetting, Filtering, and Subsampling & 
Selection of regions within an image. Thresholding of pixel values. \\ \hline
Change Detection and Comparison & 
Spatial queries to compare segmented nuclei and features within and across images. \\ \hline 
\end{tabular}
\end{footnotesize}
\label{tab:op-categories}
\vspace*{-7ex}
\end{center}
\end{table}

%
Our work examines the performance impact of different data access and processing 
patterns on application operations on CPUs, GPUs, and MICs.  
The main contributions of the paper can be summarized as follows: (1) 
We define the data access and computation patterns of operations in 
the object segmentation and feature computation categories for a microscopy 
image analysis application. (2)
We systematically evaluate the performance of the operations on modern CPUs, 
GPUs, and MIC systems. (3) The results show that 
the data access pattern and parallelization strategy employed by the operations
strongly affect their performance. While the performance on a MIC of operations that 
perform regular data
access is comparable or sometimes better than that on a GPU. GPUs are
significantly more efficient than MICs for operations and algorithms that
irregularly access data. This is a result of the low performance of the latter
when it comes to random data access. 
Coordinated execution on MICs and CPUs using a performance aware task
scheduling strategy improves about 1.29$\times$ over a first-come-first-served 
strategy. The example application attained an efficiency of 84\% in an
execution with of 192 nodes (3072 CPU cores and 192 MICs).

\section{Example Application and Core Operations} \label{sec:app}

In this section we provide a brief overview of microscopy image 
analysis as our example application. Presently our work  is focused on the development of 
operations in the object segmentation and feature computation categories, since these are 
the most expensive categories (or stages) in this application. We describe the operations 
in these stages, and present the operations' data access and processing patterns. 
\subsection{Microscopy Image Analysis}
Along with advances in slide scanners, digitization of whole slide tissues, extracted 
from humans or animals, has become more feasible and facilitated the utility of whole 
slide tissue specimens in research as well as clinical settings.  Morphological changes 
in tissues at the cellular and sub-cellular scales provide valuable information about 
tumors, complement genomic and clinical information, and can lead to a better 
understanding of tumor biology and clinical outcome~\cite{ieee-insilico}. 

Use of whole slide tissue images (WSIs) in large scale studies, involving thousands of 
high resolution images, is a challenging problem because of computational requirements 
and the large sizes of WSIs. 
%
The segmentation and feature computation stages may operate on images of 100K$\times$100K 
pixels in resolution and may identify about millions of micro-anatomic objects in an image. 
Cells and nuclei are detected and outlined during the segmentation stage. This stage applies 
a cascade of operations that include pixel value thresholds and morphological 
reconstruction to identify candidate objects, fill holes to remove holes inside objects, 
area thresholding to filter out objects that are not of interest. Distance transform
and watershed operations are applied to separate objects that overlap.  
The feature computation stage calculates a set of spatial and texture properties per
object that include pixel and gradient statistics, and edge and morphometry features. 
The next section describes the set of core operations in these two stages and 
their data and processing patterns.
\subsection{Description of Core Operations} \label{sec:core-opts}
The set of core operations in the segmentation and feature computation 
stages is presented in Table~\ref{tab:core-ops}.
These operations are categorized according to the computation stage in which
they are used (segmentation or feature computation), data access pattern, 
computation intensity, and the type of parallelism employed for speeding up 
computation. 

\begin{table*}
\begin{center}
\caption{Core operations in segmentation and feature computation phases from microscopy image analysis. IWPP stands for Irregular Wavefront Propagation Pattern.}
\begin{footnotesize}
\begin{tabular}{l l l l l}
\hline
Operations           					& Description					& Data Access Pattern 		& Computation	& Parallelism 		\\ \hline \hline
\multicolumn{5}{c}{Segmentation Phase} 																		\\ \hline
\multirow{2}{*}{Covert RGB to grayscale}		& Covert a color RGB image into			& Regular, multi-channel	& Moderate	& Data			\\ 
							&  grayscale intensity image 			& local				& 		&			\\ \hline
Morphological Open             				& Opening removes small objects and   		& Regular, neighborhood 	& Low        	& Data			\\ 
 					  		& fills small holes in foreground 		& (13x13 disk)			&               &			\\ \hline
Morphological 						& Flood-fill a marker image that is limited by	& Irregular, neighborhood 	& Low        	& IWPP			\\ 
Reconstruction~\cite{Vincent93morphologicalgrayscale}   & a mask image.					& (4-/8-connected)		&               &			\\ \hline
Area Threshold           				& Remove objects that are not within an area range& Mixed, neighborhood 	& Low     	& Reduction		\\ \hline
FillHolles              				& Fill holes in an image objects using a flood-fill 	& Irregular, neighborhood& Low		& IWPP			\\ 
 					  		& in the background pixels starting at selected points	& (4-/8-connected)	&       	&			\\ \hline
\multirow{2}{*}{Distance Transform} & Computes the distance to the closest background  & Irregular, neighborhood& Moderate 	& IWPP			\\ 
 					  		& pixel for each foreground pixel		&	(8-connected)		&               &			\\ \hline
Connected Components            			& Label with the same value pixels in components& Irregular, global		& Low 		& Union-find		\\ 
Labeling						& (objects) from an input binary image		&				&  		&			\\ \hline \hline
\multicolumn{5}{c}{Feature Computation Phase} 																	\\ \hline
\multirow{2}{*}{Color Deconvolution~\cite{Ruifrok_2001}}& Used for separation of multi-stained		& Regular, multi-channel 	& Moderate	& Data			\\ 
							& biological images into different channels	& local				&		&			\\ \hline
\multirow{2}{*}{Pixel Statistics} 			& Compute vector of statistics (mean, median, 	& Regular, access a set		& High		& Object 		\\ 
							& max, etc) for each object in the input image	& of bounding-boxed areas	&   		&			\\ \hline
\multirow{2}{*}{Gradient Statistics}			& Calculates magnitude of image gradient in x,y	& Regular, neighborhood 	& High		& Object		\\ 
							& and derive same per object features		& and bounding-boxed areas	&		&			\\ \hline	
\multirow{2}{*}{Sobel Edge} 				& Compute vector of statistics (mean, median, 	& Regular, access a set		& High		& Object		\\ 
							& max, etc) for each object in the input image	& of bounding-boxed areas	&   		&			\\

\hline
\end{tabular}
\end{footnotesize}
\label{tab:core-ops}
\vspace*{-6ex}
\end{center}
\end{table*}

The operations in the segmentation stage carry out computations on elements from the
input data domain (pixels in the case of an image), while those in the feature
computation stage additionally perform computations associated with objects. In 
regards to data access patterns, the core operations may be first classified
as: 1)~\emph{regular operations} that access data in contiguous regions
such as data scans; or 2)~\emph{irregular operations} in which data
elements to be processed are irregularly distributed or accessed in the data
domain. For some operations, data elements to be processed are only known
during execution as runtime dependencies are resolved as a result of the
computation. Examples of such operations include those that perform
flood-fill and irregular wave front propagations.  

Data accessed in the computation of a given data element may be: 
1)~{\em local} for cases in which the computation of a data element depends only on its value; 
2)~{\em multi-channel local}, which is a variant of the former in operations that 
independently access data elements with the same index across multiple layers 
of the domain (e.g., multiple channels in an image); 
3)~{\em within a neighborhood}, which refers to cases when an operation performs 
computations on data elements in a spatial and/or temporal neighborhood.  
The neighborhood is often defined using structure elements such as
4-/8-connected components or discs; and 4)~{\em areas in a bounding-box}, which 
are used in the feature computation phase in operations on objects that are 
defined within minimum bounding boxes. 

Parallel execution patterns exhibited by the core operations are diverse:
1)~Data parallelism; 2)~Object parallelism; 3)~MapReduce~\cite{dean04mapreduce}
or generalized reduction; 4)~Irregular wavefront propagation pattern
(IWPP)~\cite{Teodoro:2013:Parco,Teodoro-2012-Morph};
5)~Union-find~\cite{Tarjan:1975:EGB:321879.321884}.  The \emph{data parallel}
operations are those that concurrently and independently process elements of the
data domain. The \emph{Object parallelism} exists in operations that process 
multiple objects concurrently. Moreover, a
\emph{MapReduce-style pattern} is also used in the ``Area Threshold" operation.
This operation maps elements from the input data according to their values
(labels) before a reduction is performed to count the number of elements with
the same label value (area of components). The area is then used to filter out
components that are not within the desired size range. 

The \emph{IWPP} pattern is characterized by independent wavefronts that start in one or
more elements of the domain. The structure of the waves is dynamic, irregular,
data dependent, and only known at runtime as expansions are computed. The
elements forming the front of the waves (active elements) work as sources of
propagations to their neighbors, and only active elements are the ones that
contribute to the output. Therefore, the efficient execution of this pattern
relies on using a container structure, e.g., a queue or a set, to maintain the
active elements and avoid computing areas of the data domain that are not
effectively contributing to the output. This pattern is presented in
Algorithm~\ref{alg:genericRecon}. A set of (active) elements from a
multi-dimensional grid space ($D$) is selected to compose the wavefront ($S$).
During the propagations, an element ($e_i$) from S is extracted and the
algorithm tries to propagate $e_i$ value to its neighbors ($N_G$) in a
structure $G$. If a propagation condition between the active element and each
of the neighbors ($e_j \in Q$) is evaluated true, the propagation occurs and
that neighbor receiving the propagation is added to the set of active elements
($S$). This process occurs until the set $S$ is not empty. The parallelization
of this pattern heavily relies on an efficient parallel container to store the
wavefront elements. In the parallel version of IWPP multiple elements from the
active set may be computed in parallel as long as race conditions that may
arise due parallel propagations that update the same element $e_j$ in the grid are
avoided. Applications that use this pattern, in addition to our core operations,
include: Watershed, Euclidean
skeletons, skeletons by influence zones, Delaunay triangulations, Gabriel
graphs and relative neighborhood graphs.
\begin{algorithm}
\begin{small}
\caption{Irregular Wavefront Propagation Pattern}
\label{alg:genericRecon}
\begin{algorithmic}[1]
\STATE $D \leftarrow$ data elements in a multi-dimensional space
\STATE \{{\bf Initialization Phase}\}
\STATE $S \leftarrow$ subset active elements from $D$
\STATE \{{\bf Wavefront Propagation Phase}\}
\WHILE{$S \neq \emptyset$}
	\STATE Extract $e_i$ from $S$
	\STATE $Q \leftarrow$ $N_G(e_i)$
	\WHILE{$Q \neq \emptyset$}
		\STATE Extract $e_j$ from $Q$
		\IF{$PropagationCondition$($D(e_i)$,$D(e_j)$) $=$ true}
			\STATE $D(e_j) \leftarrow$ $Update$($D(e_i)$)
			\STATE Insert $e_j$ into $S$
		\ENDIF
	\ENDWHILE
\ENDWHILE 
\end{algorithmic}
\end{small}
\end{algorithm}

The \emph{union-find pattern}~\cite{Tarjan:1975:EGB:321879.321884} is used for
manipulating disjoint-set data structures and is made up of three
operations: 1)~Find: determines the set in which a component is stored;
2)~Union: merges two subsets into a single set; and 3)~MakeSet: creates an
elementary set containing a single element.  This is the processing structure of 
the connected components labeling (CCL) operation in our implementation. 
The CCL first creates a forest in which each element (pixel) from
the input data is an independent tree. It iteratively merges trees from
adjacent elements in the data domain such that one tree becomes a branch in another
tree. The condition for merging trees is that the neighbor elements must be
foreground pixels in the original image. When merging two trees (Union), the label
values of the root of the two trees are compared, and the root with the smaller
value is selected as the root of the merged tree.  After this
process is carried out for all pixels, each connected component is assigned to
a single tree, and the labeled output can be computed by flattening the trees
and reading labels.


\section{Implementation of Core Operations}



\subsection{Architectures and Programming Models} \label{sec:arch}
We have implemented the operations listed in 
Table~\ref{tab:core-ops} for the new Intel Xeon Phi (MIC), CPUs, and GPUs. 
We briefly describe the MIC architecture only, because of space constraints. 
The MIC used in the experimental evaluation (SP10P) is built using 61 light-weight x86 
cores clocked at 1090 MHz, with cache coherency among all cores that are 
connected using a ring network. The cores process instructions in-order, support 
a four-way simultaneous multi-threading (SMT), and execute 512-bit wide SIMD 
vector instructions. The MIC is equipped with 8GB of GDDR5 DRAM with theoretical
peak bandwidth of 352GB/s. 

The programming tools and languages employed for code development for a MIC are
the same as those used for CPUs. This is a significant advantage as compared
to GPUs that alleviates code migration overhead for the co-processor. The MIC
supports several parallel programming languages and models, such as OpenMP,
POSIX Threads, and Intel Cilk Plus. In this work, we have implemented our
operations using OpenMP on the MIC and CPU; we used 
CUDA\footnote{http://nvidia.com/cuda/.} for the GPU implementations.
The MIC supports two execution modes:
native and offload. In the native mode the application runs entirely within the
co-processor. This is possible because MICs run a specialized Linux kernel that
provides the necessary services and interfaces to applications. The offload
mode allows for the CPU to execute regions of the application code with the
co-processor. These regions are defined using pragma tags and include
directives for transferring data. The offload mode also supports conditional offload
directives, which the application developer may use to decide at runtime whether a region
should be offloaded to the coprocessor or should be executed on the CPU.
This feature is used in our dynamic task assignment strategy for application execution
using the CPU and the MIC cooperatively.  
\subsection{Parallel Implementation of Operations} 
\label{sec:opt-impl}
We present the details of the MIC and GPU implementations only, because
the CPU runs the same code as the MIC.  The \emph{Data parallel} operations
were trivial to implement, since computing threads may be assigned for
independent computation of elements from the input data domain. However, we had
to analyze the results of the auto vectorization performed by the compiler for the MIC,
because it was not be able to vectorize some of the loops when complex pointer
manipulations were used. In these cases, however, we annotated the code with
(\#pragma simd) to guide the vectorization when appropriate.

The parallelization of operations that use the \emph{IWPP pattern} heavily
relies on the use of parallel containers to store the wavefront elements. The
parallel computation of elements in the wavefront requires those elements be
atomically updated, since multiple elements may concurrently update a third
element $e_j$. In order to implement this operation in CUDA, we developed a
complex hierarchical parallel queue to store wavefront
elements~\cite{Teodoro:2013:Parco}. This parallel queue exploits the multiple
GPU memory levels and is implemented in a thread block basis, such that each
block of threads has an independent instance of the queue to avoid
synchronization among blocks. The implementation of the IWPP on the MIC was much
simpler. The standard C++ queue container used in the sequential
version of IWPP is also available with the MIC coprocessor. Thus, we
instantiated one copy of this container per computing thread, which independently
carries out propagations of a subset of wavefront elements. In both cases, atomic
operations were used to update memory during a propagation to avoid race
conditions and, as a consequence, the MIC vectorization was not possible since
vector atomic instructions are not supported.

The \emph{MapReduce-style pattern}, or reduction, is employed in object
area calculations. The MIC and GPU implementations use a vector with an entry per
object to accumulate its area, and threads concurrently scan pixels in the
input data domain to atomically increment the corresponding entry in the
reduction vector. Because the number of objects may be very high, it is not
feasible to create a copy of this vector per thread and eliminate the use of
atomic instructions. 

In the \emph{Union-find pattern} a forest is created in the input image, such
that each pixel stores its neighbor parent pixel or itself when it is a tree
root. For the parallelization of this pattern, we divided the input data into
tiles that may be independently processed in parallel. A second phase was then
executed to merge trees that cross tile boundaries. The MIC implementation
assigns a single tile per thread and avoids the use of atomic instructions in
the first phase. The GPU implementation, on the other hand, computes each tile
using a thread block. Since threads computing a tile are in the same block,
they can take advantage of fast shared-memory atomic instructions. The second
phase of Union-find was implemented similarly for the MIC and the GPU. It uses
atomic updates to guarantee consistency during tree merges across tile
boundaries.

Finally, the operations with \emph{Object Parallelism} can be independently
carried out for each segmented object. Therefore, a single thread in
the MIC or a block of threads in the GPU is assigned for the computation of
features related to each object. All of the operations with this type of
parallelism were fully vectorized.

\section{Cooperative Execution on Clusters of Accelerators} \label{sec:dist_mic}
%
The strategy 
for execution of a pipeline of segmentation and feature computation stages on a cluster 
system is based on a Manager-Worker model
that combines a bag-of-tasks style execution with coarse-grain dataflow and
makes use of function variants.  A function variant represents multiple
implementations of a function with the same signature. In our case, the
function variant of an operation is the CPU, GPU, and MIC implementations. One
Manager process is instantiated on the head node. Each computation node is
designated as a Worker. The Manager creates tasks of the form (input image
tile, processing stage), where {\em processing stage} is either the
segmentation stage or the feature computation.  Each of these tasks is referred
to as a stage task. The Manager also builds the dependencies between stage task
instances to enforce correct execution. The stage tasks are scheduled to the
Workers using a demand-driven approach.  A Worker may ask for multiple tasks
from the Manager in order to keep all the computing devices on a node busy. 

A local Worker Resource Manager (WRM) on each computation node controls the CPU
cores and co-processors (GPUs or MICs) used by a Worker. When the Worker
receives a stage task, the WRM instantiates the operations in the stage task.
It dynamically creates operation tasks, represented by tuples (input data,
operation), and schedules them for execution as it resolves the dependencies
between the operations -- note that the segmentation and feature computation
stages consist of pipelines of operations; hence there are dependencies between
the operations.  The set of stage tasks assigned to a Worker may create many
operation tasks. The operations may have different performance characteristics
on different computing devices.  In order to account for this variability, a
task scheduling strategy, called Performance Aware Task Scheduling (PATS), was
employed in our
implementation~\cite{Teodoro-IPDPS2012,DBLP:conf/ipps/TeodoroPKKCPKS13}.  PATS
assigns tasks to CPU cores or co-processors based on an estimate of each task's
co-processor speedup and on the computational loads of the co-processors and
CPUs. When an accelerator requests a task, PATS assigns the tasks with higher
speedup to this processor. If the device available for computation is a CPU,
the task to attain lower speedup on the accelerator is chosen.  We refer
reader to~\cite{Teodoro-IPDPS2012,DBLP:conf/ipps/TeodoroPKKCPKS13} for a more
detailed description of the PATS implementation.  PATS also implements
optimizations such as data reuse and overlap of data copy and computations to
further reduce data processing overheads.

\section{Experimental Evaluation} \label{sec:results}
We carried out the experimental evaluation using a distributed memory Linux
cluster, called Stampede\footnote{https://www.xsede.org/tacc-stampede}.  Each
compute node has dual socket Intel Xeon E5-2680 processors, an Intel Xeon Phi
SE10P co-processor, and 32GB of main memory. We also used a node equipped with
a single NVIDIA K20 GPUs.  The nodes are connected using Mellanox FDR
InfiniBand switches. The codes were compiled using Intel Compiler 13.1 with
``-O3'' flag.  Since we execute the codes on the MIC using the {\em offload
mode}, a computing core is reserved to run the offload daemon, and a maximum of
240 computing cores are launched. The images for the experiments were collected
from brain tumor studies~\cite{ieee-insilico}. Each image is divided into
4K$\times$4K tiles which are processed concurrently on the cluster system.

\subsection{Scalability of Operations on MIC} \label{sec:opts-scalability}
This section evaluates the performance and scalability of the operations on the
MIC. This analysis also considers the effects of thread affinity that
determines the mapping of computing threads to computing cores.  We examined
three affinity strategies: compact, balanced, and scatter. {\em Compact}
assigns threads to the next free thread context $n+1$, i.e., all four contexts
in a physical core are used before threads are placed in the contexts of
another core. {\em Balanced} allocates threads to new computing cores before
contexts in the same core are used.  Threads are balanced among computing cores
and subsequent thread IDs are assigned to neighbor contexts or cores.  {\em
Scatter} allocates threads in a balanced way, like the balanced strategy, but
it sets thread IDs such that neighbor threads are placed in different computing
cores.  We selected two operations with different data access and computation
intensities for the experiments: (1)~Morphological Open has a regular data
access pattern and low computation intensity;  (2)~Distance Transform performs
irregular data access and has a moderate computation intensity.  OpenMP static
loop scheduling was used for execution, because the dynamic version resulted in
lower performance.

\begin{figure}[htb!]
\begin{center}
	\subfigure[]{\label{fig:speedup-open}
        \includegraphics[width=0.46\textwidth]{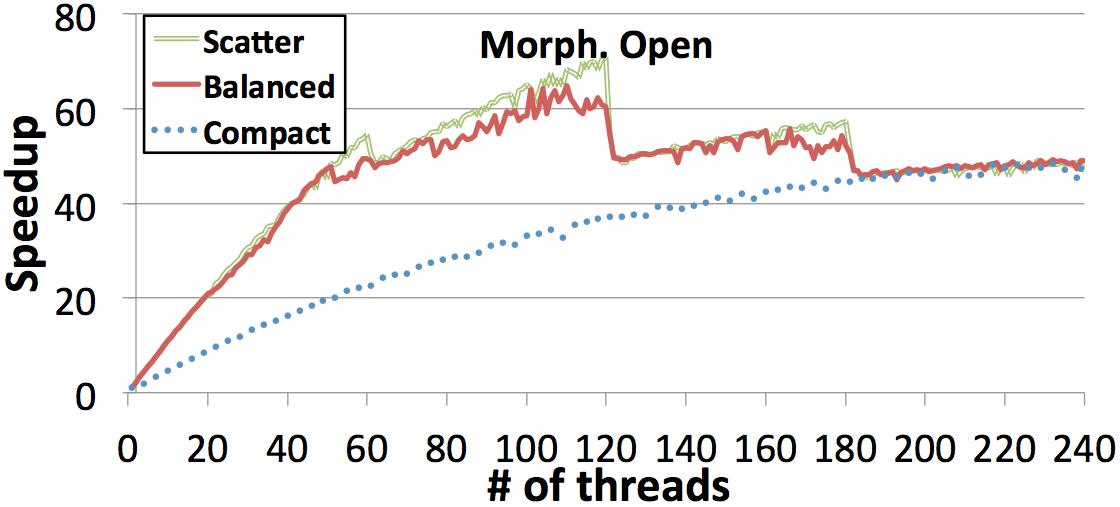}}
	\subfigure[]{\label{fig:speedup-dist}
        \includegraphics[width=0.46\textwidth]{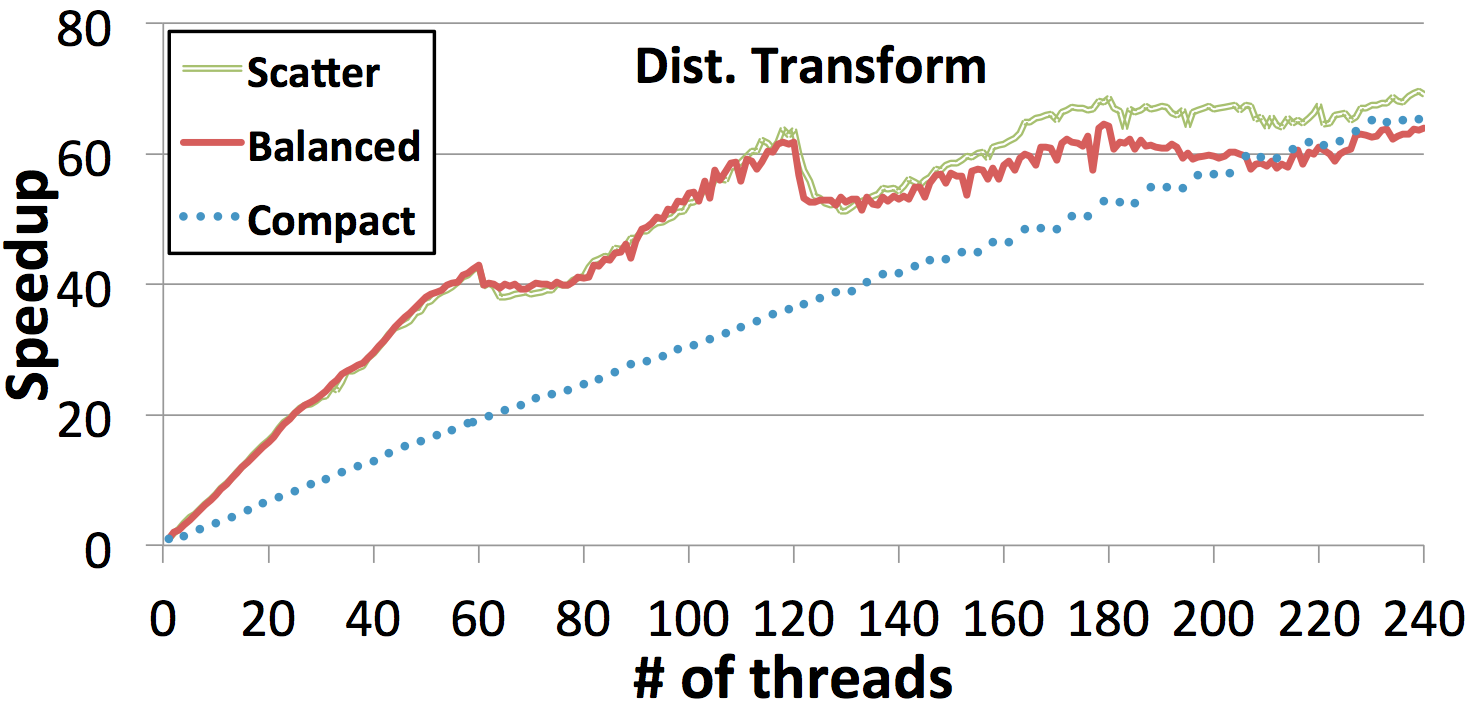}}
\vspace*{-2ex}
\caption{Evaluation of scalability with respect to thread affinity type for selected operations on the MIC.}
\vspace*{-3ex}
\label{fig:opts-scalability}
\end{center}
\end{figure}

The scalability results with respect to thread affinity are presented
in Figure~\ref{fig:opts-scalability}. As is shown, there is a great variability
in speedups with different operations and different thread affinity strategies.
Morphological Open and Distance Transform achieved the best
performances when 120 and 240 threads were used, respectively
(Figures~\ref{fig:speedup-open} and~\ref{fig:speedup-dist}). The graphs show
that peaks in performance are reached when the number of threads is a multiple of
the number of computing cores, i.e., 60, 120, 180, and 240 threads. In these
cases, the number of threads allocated per computing core is the same; 
hence, computational work is better balanced among the physical cores. 

The performance of the Morphological Open operation scales until 120 threads are 
employed. Its performance significantly degrades when more threads are
used. This behavior is a consequence of the MICs performance with memory
intensive applications.  As reported
in~\cite{McCalpin1995,DBLP:conf/ipps/SauleC12}, the maximum  
memory bandwidth on the MIC is reached, measured using the STREAM 
benchmark~\cite{McCalpin1995} (regular data access), when one or two threads 
are instantiated per computing core (a total of 60 or 120 threads).  When the number of threads
increases to 180 and 240, there is a reduction in memory throughput due to 
congestion on the memory subsystem. Since Morphological Open is memory
bound, this property of the MIC is a limiting factor on the performance
of the operation when 120 and more threads are executed.  

The scalability of Distance Transform is presented in
Figure~\ref{fig:speedup-dist}.  This operation fully benefited from the MIC's
4-way hyperthreading and attained the best performance with 240 threads. 
Memory bandwidth also plays an important role in the scalability of this operation.
However, this operations performs irregular access to data. 
Since
random memory bandwidth is not typically included in devices specifications, we
created a micro-benchmark to the MIC's 
performance with random
data access in order to better understand the operation’s
performance.
This benchmark consists of a program that randomly reads and writes
elements from/to a matrix in parallel. The positions to be accessed in this
matrix are stored into a secondary vector of indices, which is equally
initialized in the CPU for all the devices.
We observed that bandwidth attained by the MIC
with random data access increases until the number of threads is 240.
We conclude that this 
is the reason for 
the observed performance behavior of the Distance Transform operation. 

The scatter and balanced thread affinity strategies achieve similar
performance, but the compact strategy fails to attain good performance with the
Morphological Open operation. This is because the compact strategy uses 
all the 60 physical cores only when close to 240 threads are instantiated. 
On the other hand, 
Morphological Open achieves best performance with 120 threads. 
With this many threads the compact affinity uses only 30 physical cores.
\subsection{Performance Impact of Vectorization on MIC}
\label{sec:vectorization}
This section analyzes the impact of using the Intel Xeon Phi SIMD capabilities.
For this evaluation, we used the Gradient Stats operation, which makes full use of SIMD
instructions to manipulate single-precision floating-point data. The scatter
affinity strategy is used in the experiments because it was more
efficient. Gradient Stats achieved speedups of
16.1$\times$ and 39.9$\times$, respectively, with the non-vectorized and
vectorized versions, as compared to the single-threaded vectorized execution.
The performance gains with vectorization are higher with lower number of
threads -- 4.1$\times$ for the single-threaded configuration. The performance 
gap between the two versions is reduced as the number of threads
increases because of the better scalability of the non-vectorized version.
At best, vectorization results in an improvement of 2.47$\times$ on top of the
non-vectorized version.
\subsection{Comparative Performance of MIC, GPU, and CPU} \label{sec:comp-performance}
This section evaluates the performance of the operations on the MIC, GPU, and
CPU. The speedup values were calculated using the
single core CPU execution as the baseline. While the CPU and MIC executables
were generated from the same C++ source code annotated with OpenMP, the GPU 
programs were implemented using CUDA. The same parallelization strategy 
was employed in all of the implementations of an operation.
\begin{figure*}[htb!]
\begin{center}
\includegraphics[width=\textwidth]{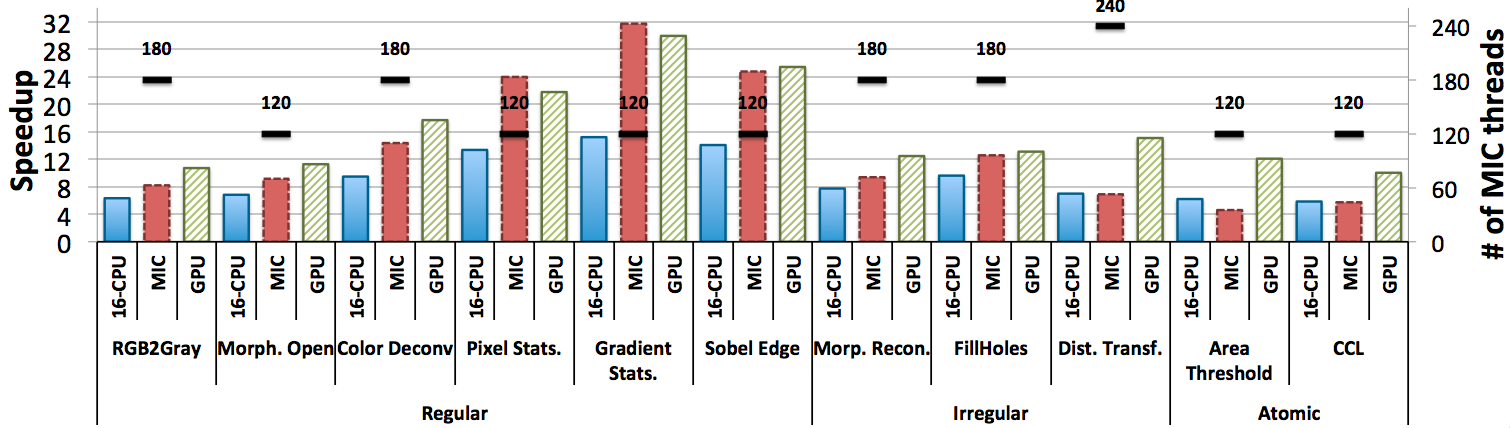}
\vspace*{-5ex}
\caption{Speedups achieved by operations on the CPU, MIC, and GPU, using the single core CPU version 
as a reference. The number above each dash refers to the number of threads that lead to the best performance 
on the MIC.}
\vspace*{-4ex}
\label{fig:comp-performance}
\end{center}
\end{figure*}

The overall performance of the operations on different processors is presented
in Figure~\ref{fig:comp-performance}. There is high variability in
the speedup attained by each of the operations even when a single device is
considered. In addition, the relative performance on the CPU, MIC, and GPU 
varies among the operations, which suggests that different computing devices are more
appropriate for specific operations.  In order to understand the reasons for 
the performance variations, we
divided the operations into three disjoint groups that internally have
similarities regarding the memory access pattern and execution strategy. The groups
are: (1)~Operations with regular data access: RGB2Gray, Morphological Open,
Color Deconvolution, Pixel Stats, Gradient Stats, and Sobel Edge; (2)~Operations with
irregular data access: Morphological Reconstruction, FillHoles, and Distance
Transform; (3)~Operations that heavily rely on the use of atomic functions,
which include the Area Threshold and Connected Component Labeling (CCL). 
To understand the performance of these operations, we measured their
computation intensity and correlated it with each device's capabilities using
the notions of the Roofline model~\cite{Williams:2009:RIV:1498765.1498785}. 
\subsubsection{Regular Operations}  
The peak memory bandwidth and computation
power of the processors are important to analyze the operations performance on
each of them. The memory bandwidth with regular data access was measured using
the STREAM benchmark~\cite{McCalpin1995} in which the K20 GPU, the CPU, and the
MIC reached peak throughputs of 148GB/s, 78GB/s (combined for the two CPUs),
and 160GB/s, respectively, with a single thread per core. Increasing the number of
threads per core with the MIC results in a reduction of the bandwidth. Moreover, 
while the K20 GPU and the MIC are expected to deliver peak double
precision performance of about 1~TFLOPS, the 2 CPUs together achieve 345~MFLOPS.

The Morphological Open, RGB2Gray and Color Deconvolution operations are memory bound
operations with low arithmetic-instruction-to-byte ratio. As presented in
Figure~\ref{fig:comp-performance}, their performance on the GPU is about
1.25$\times$ higher than that on the MIC. Furthermore, the CPU scalability with
this operations is low, because the memory bus is rapidly saturated. The
improvements of the GPU on top of the CPU are consistent with their differences
in memory bandwidth. The Color Deconvolution operation attains better raw
speedups than other operations due to its higher computation intensity. While
Morphological Open attains maximum performance with 120 threads because of its
ability of reusing cached data (neighborhood in computation of different
elements may overlap), the other two operations use 180 threads in order to
hide the memory access latency. The remaining of the regular operations (Pixel
Stats, Gradient Stats, and Sobel Edge) are compute bound due to their higher
computation intensity. These operations achieve better
scalability with all the devices. The performances of the GPU and MIC are 
similar with improvements of about 1.9$\times$ on
top of the multicore CPU execution because of their higher computing
capabilities. This set of compute intensive operations obtained the best
performance with the MIC using 120 threads. Using more threads does not 
improve performance because the MIC threads can launch a vector instruction
each two cycles, and compute intensive operations should maximize the hardware
utilization with a 2-way hyperthreading.
\subsubsection{Irregular Operations}
The operations with irregular data access patterns are Morphological  Reconstruction,
FillHoles, and Distance Transform. This set of operations strongly relies on the 
device performance to execute irregular (random) accesses to data. 
We used the same micro-benchmark described in Section~\ref{sec:opts-scalability}
to measure each of the systems' throughput in this
context. 

\begin{table}
\caption{Device Bandwidth with Random Data Accesses (MB/s).}
\vspace*{-4ex}
\begin{center}
\begin{tabular}{l l l l}
\hline
   	& CPU	& MIC	&	GPU 	\\ \hline \hline
Reading	& 305	& 399	&	895	\\ \hline
Writing	& 74	& 16	&     	126	\\ \hline
\end{tabular}
\end{center}
\label{tab:random-bandwidth}
\vspace*{-4ex}
\end{table}

The results are presented in
Table~\ref{tab:random-bandwidth}. The experiments were carried out by executing 10
million random reading or writing operations in a 4K$\times$4K matrix of
integer data elements. As shown, the bandwidth attained by the processors is
much lower than those with regular data access. The GPU significantly
outperforms the other devices. The random writing bandwidth of the MIC
processors is notably poor. This is in fact expected because this processor
needs to maintain cache consistency among its many cores, which will result in
a high data traffic and competition in its ring bus connecting caches. Due the
low bandwidth attained by all the processors, all of our irregular
operations are necessarily memory bound.

As presented in Figure~\ref{fig:comp-performance}, the Distance Transform  operation
on the GPU is about 2$\times$ faster than on the MIC, whereas the MIC performance
is not better than that of the CPU.  This operation performs only irregular data
access in all phases of its execution, and the differences in the random data
access performances of the devices are crucial to its performance. 

The other two operations (Morphological Reconstruction and Fill Holes) in this category have a
mixed data access patterns. These operations are based on the irregular wave
front propagation pattern, and their most efficient execution is carried out
with an initial regular propagation phase using raster/anti-raster scans,
before the algorithm migrates to the second phase that irregularly access
data and uses a queue to execute the propagations. Since the first phase of
these operations is regular, it may be efficiently executed on the MIC. In the
MIC execution, the algorithm will iterate several times over the data using the
regular (initial) phase, before it moves to the irregular queue based phase. 
The MIC execution will only migrate to the irregular pattern
after most of the propagations are resolved, which reduces the amount of time
spent in the irregular phase. Hence, the
performance gains on the GPU as compared to those on the MIC are smaller for both
operations: about 1.33$\times$. We want to highlight that the same tuning
regarding the appropriate moment to migrate from the regular to the irregular
phase is also performed with the GPU version.
\subsubsection{Operations that Rely on Atomic Instructions} 
The Area Threshold and CCL operations heavily rely on the use of atomic add
instructions to execute a reduction. Because the use of atomic
instructions is critical for several applications and computation patterns, we
analyze the performance of the evaluated devices with regard to execution of atomic
instructions and its implications to the Area Threshold and CCL operations. To
carry out this evaluation, we created a micro-benchmark in which computing
threads concurrently execute atomic add instructions in two scenarios:
(1)~using a single variable that is updated by all threads and (2)~an array
indexed with the thread identifier. The first configuration intends to measure
the worst case performance in which all threads try to update the same memory
address, whereas the second case assesses 
performance with threads updating disjoint memory addresses.

\begin{table}
\caption{Device Throughput with Atomic Adds (Millions/sec).}
\vspace*{-4ex}
\begin{center}
\begin{tabular}{l l l l}
\hline
   		& CPU	& MIC		&	GPU 	\\ \hline \hline
Single Variable	& 134	& 55		&	693	\\ \hline
Array		& 2,200	& 906		&	38,630  \\ \hline
\end{tabular}
\end{center}
\label{tab:atomic}
\vspace*{-4ex}
\end{table}

The results are presented in Table~\ref{tab:atomic}.  The GPU
once again attained the best performance, and it is at least 5$\times$ faster
than the other processors in both scenarios. The reduction in the GPU
throughput from the configuration with an array to the single variable,
however, is the highest among the processors evaluated. This drastic reduction
in performance occurs because a GPU thread warp executes in a SIMD way and, hence, 
the atomic instructions launched by all threads in a warp will be
serialized. In addition, the GPU
launches a larger number of threads. This results in higher levels of
concurrency and contention for atomic instructions. 

The CPU is about 2.4$\times$ faster than the MIC in both scenarios. The MIC
is equipped with simpler computing core and typically relies on the
use of vectorized operations to achieve high performance. However, it lacks
support for vector atomic instructions, which poses a serious limitation.  The
introduction of atomic vector instructions, such as those proposed by Kumar et.
al.~\cite{4556746} for other multiprocessors, could significantly improve the
MIC performance. Because the Area Threshold and CCL operations greatly depend
on atomic instructions, they attained better performance on the GPU. In both
cases, the execution on the CPU is more efficient than on the MIC
(Figure~\ref{fig:comp-performance}).

\subsection{Multi-node Execution using CPUs and MICs}  \label{sec:cpu-gpu-base}

This section evaluates application performance in using CPUs and MICs
cooperatively on a distributed memory cluster. The example application is built
from the core operations and is implemented as a hierarchical pipeline in which
the first level is composed of segmentation and feature computation stages, and
each of these stages is created as a pipeline of operations as described in
Section~\ref{sec:dist_mic}. We evaluated four versions of the application:
(1)~CPU-only refers to the multi-core CPU version that uses all CPU cores
available; (2)~MIC-only uses a MIC per node to perform computations;
(3)~CPU-MIC FCFS uses all CPU cores and the MIC in coordination and distributes
tasks to processors in each node using FCFS (First-Come, First-Served) fashion;
(4)~CPU-MIC PATS also uses all CPU cores and the MIC in coordination, but the
tasks are scheduled to devices based on the expected speedup of each task on
each device.  The speedup estimates are those presented
in Figure~\ref{fig:comp-performance}.

\begin{figure}[htb!]
\begin{center}
        \includegraphics[width=0.49\textwidth]{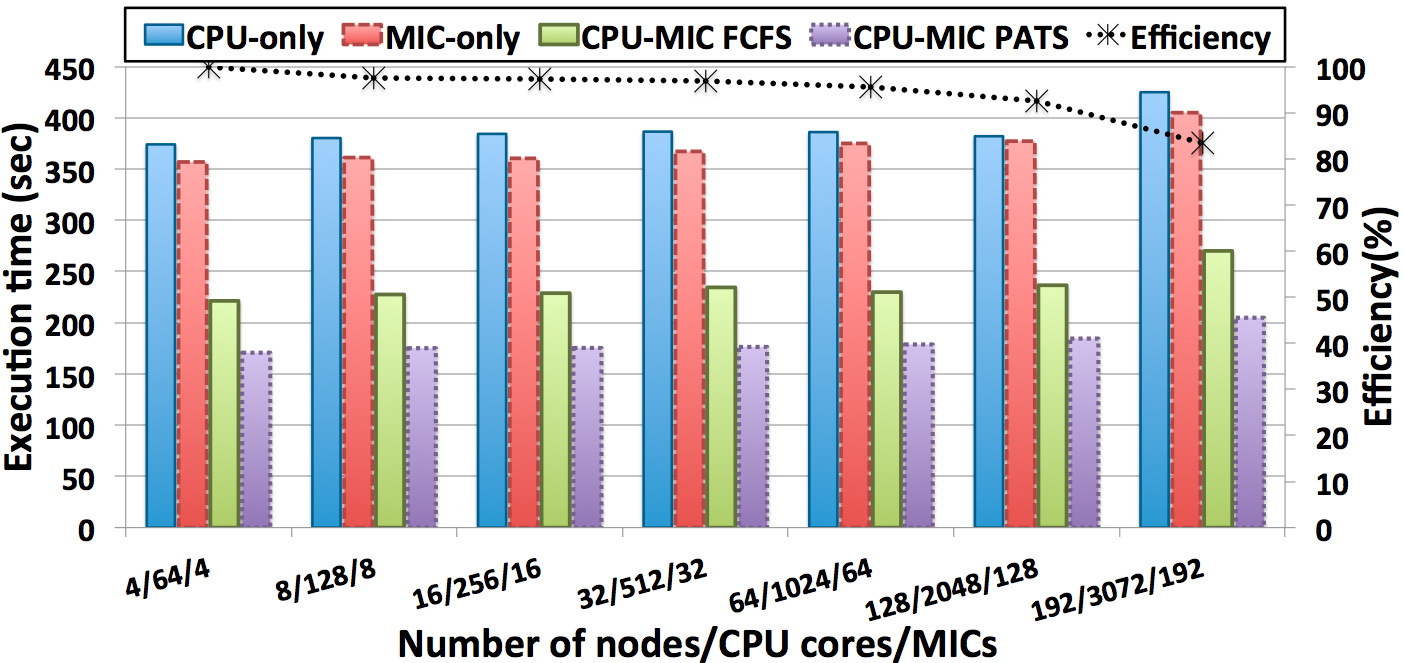}
\vspace*{-2ex}
\caption{Multi-node weak scaling evaluation: dataset size and the number of nodes increase proportionally. }
\vspace*{-4ex}
\label{fig:weak-scale-general}
\end{center}
\end{figure}

The weak scaling evaluation in which the dataset size and the number of nodes
increase proportionally is presented in Figure~\ref{fig:weak-scale-general}.
The experiments with 172 nodes used an input dataset with 68,284 4K$\times$4K
image tiles (3.27TB of uncompressed data). All versions of the application
scaled well as the number of nodes is increased from 4 to 192. The MIC-only
execution was slightly faster than the multi-core CPU-only version.  The
cooperative CPU-MIC executions attained improvements of up to 2.06$\times$ on
top of the MIC-only version. The execution using PATS is 1.29$\times$ faster
than using FCFS. This is a result of PATS being able to map tasks to the more
appropriate devices for execution. The efficiency of the fastest CPU-MIC PATS
version is about 84\%, when 192 computing nodes are used. The main factor
limiting performance is the increasing cost of reading the input image tiles
concurrently from disk as the number of nodes (and processes) grows.

\section{Related Work} \label{sec:related}
Efficient utilization of computing systems with co-processors 
requires the implementation of efficient and scalable application 
computing $kernels$, coordination of assignment of work to co-processors 
and CPUs, minimization of communication, and overlapping of communication 
and computation. Mars~\cite{mars} and Merge~\cite{merge} are designed to 
enable efficient execution of MapReduce computations on shared memory machines
equipped with CPUs and GPUs. Qilin~\cite{qilin09luk} implements an automated 
methodology to map computation tasks to CPUs and GPUs. 
PTask~\cite{Rossbach:2011:POS:2043556.2043579} provides OS abstractions for 
task based applications on GPU equipped systems. 
Other frameworks to support execution on distributed memory
machines with CPUs and GPUs were also
proposed~\cite{6061070,ravi2010compiler,HartleySC10,hpdc10george,6267858,Teodoro:2012:Cluster,augonnet:hal-00725477,sbac09georgeAnthill,cluster09george,Teodoro-vldb}.
DAGuE~\cite{6061070} and StarPU~\cite{augonnet:hal-00725477} support execution
of regular linear algebra applications. They express the
application tasks dependencies using a Directed Acyclic Graph (DAG) and provide
different scheduling policies, including those that prioritize execution of
tasks in the critical path.  Ravi~\cite{ravi2010compiler} and
Hartley~\cite{HartleySC10} proposed runtime techniques to auto-tune work
partitioning among CPUs and GPUs. OmpSs~\cite{6267858} supports execution 
of dataflow applications created via compilation of annotated code.

More recently, research groups have focused on applications that may
benefit from Intel's Xeon Phi
co-processor~\cite{phi-1,6569806,phi-3,phi-4,DBLP:conf/ipps/SauleC12}. 
Jo\'o et al.~\cite{phi-1} implemented a Lattice Quantum
Chromodynamics (LQCD) method
using Intel Phi processors. Linear algebra algorithms were also ported to
MIC~\cite{6569806,phi-4}. Hamidouche et
al.~\cite{Hamidouche:2013:MEE:2464996.2465445} proposed an
automated approach to perform computation offloads on remote nodes. The use of
OpenMP based parallelization on a MIC processor was evaluated in~\cite{phi-3}.
That work analyzed the overheads of creating and synchronizing threads, processor
bandwidth, and improvements with the use of vector instructions. 
Saule et al.~\cite{DBLP:conf/ipps/SauleC12} implemented optimized sparse matrix
multiplication kernels for MICs, and provided a comparison of MICs and
GPUs for this operation.
 
In our work, we perform a comparative performance evaluation of
MICs, multi-core CPUs, and GPUs using an important class of operations.
These operations employ diverse computation and data access patterns and
several parallelization strategies. The comparative performance
analysis correlates the performance of operations with co-processors
characteristics using co-processor specifications or performance measured
using micro-kernels. This evaluation provides a methodology and clues for 
application developers to understand the efficacy of co-processors for 
a class of applications and operations. We also investigate coordinated use of 
MICs and CPUs on a distributed memory machine and its impact on application 
performance. Our approach takes into account performance variability of 
operations to make smart task assignments.


\section{Conclusions} \label{sec:conclusions}
Creating
efficient applications that fully benefit from systems with co-processors 
is a challenging problem. New co-processors are being released with more 
processing and memory capacities, 
but application developers often have little information about which co-processors 
are more suitable for their applications. In this paper we provide a comparison of
CPUs, GPUs, and MICs using operations, which exhibit different data
access patterns (regular and irregular), computation intensity, and types of
parallelism, from a class of applications. An array of parallelization strategies commonly
used in several applications are studied. The experimental results show
that different types of co-processors are more appropriate for specific data
access patterns and types of parallelism, as expected. The MIC's
performance compares well with that of the GPU when regular operations and computation
patterns are used. The GPU is more efficient for those operations that perform irregular 
data access and heavily use atomic operations. A strong
performance variability exists among different operations, as a result of
their computation patterns. This variability needs to be taken into 
account to efficiently execute pipelines of operations  
using co-processors and CPUs in coordination. Our results show that 
the example application can achieve 84\% efficiency on a distributed memory 
cluster  of 3072 CPU cores and 192 MICs using a performance aware task scheduling 
strategy. 

{\bf Acknowledgments.}
{\small This work was supported in part by
HHSN261200800001E from the NCI, R24HL085343 from the
NHLBI, by R01LM011119-01 and R01LM009239 from
the NLM, and RC4MD005964 from the NIH, and CNPq.
This research was supported in part by the NSF through 
XSEDE resources provided by the XSEDE Science 
Gateways program.}


%

\balance

\bibliographystyle{IEEEtran}
{\small
\begin{scriptsize}
\bibliography{george}
\end{scriptsize}
}

\end{document}